\documentclass[a4paper]{mem}
\usepackage{natbib}
\usepackage{graphicx}
\usepackage[a4paper]{hyperref}
\idline{1}{1}
\begin{document}
   \title{Shock acceleration and gamma radiation in the intracluster medium
}

   \author{Stefano Gabici 
\thanks{E-mail: gabici@arcetri.astro.it }
}

   \offprints{S. Gabici}
\mail{Osservatorio Astrofisico di Arcetri, largo E.~Fermi 5, 50125 Firenze, Italy}

   \institute{Dipartimento di Astronomia e Scienza dello Spazio, largo E.~Fermi 5, 50125 Firenze, Italy}

   \abstract{Particle acceleration is expected to take place at shocks that form during the process of large scale structure formation. Electrons accelerated at such shocks can upscatter a small fraction of the photons in the cosmic microwave background up to the gamma ray band. Here we make predictions about the detectability of the $\gamma$--ray emission from forming clusters of galaxies with future GeV and TeV gamma ray telescopes. We also estimate the contribution of these sources to the extragalactic diffuse gamma ray background.}
   \authorrunning{S. Gabici}
   \titlerunning{Gamma rays from the intracluster medium}
   \maketitle

\section{Introduction}

Non--thermal emission from rich clusters is observed in the radio \citep{gigia} and hard X--ray \citep{fusco} bands while only upper limits for their gamma ray fluxes have been obtained from an accurate analysis of EGRET data \citep{olaf}.

Despite the fact that to date none of the nearby rich clusters has been detected, there are theoretical arguments suggesting that clusters could be sources of high energy radiation.
\citet{bbp} and \citet{vab} first understood that cosmic ray protons accelerated in the intracluster medium (ICM) would be confined there for cosmological time scales, enhancing the possibility of inelastic proton--proton collisions and consequent gamma ray production through neutral pion decay. Moreover, shocks associated with large scale structure formation might accelerate electrons up to TeV energies. These electrons can upscatter a small fraction of the photons in the cosmic microwave background radiation up to gamma ray energies through inverse Compton scattering (ICS) \citep{l&w}.
For these reasons the detection of gamma ray emission from the ICM represents one of the major goals of future gamma ray telescopes.

Here we addressed the issue of the detectability of clusters in gamma rays and of the contribution of large scale structure formation to the extragalactic diffuse gamma ray background (EDGRB).
We restrict our attention only to ICS of relativistic electrons accelerated at shocks generated during structure formation. The contribution from protons to the gamma ray flux can be of great importance, but it is also quite uncertain because, due to confinement, the amount of protons diffusively stored in the ICM depends on the whole history of the cosmic ray sources in clusters. Any contribution to the gamma ray emission from protons can only increase the fluxes derived below, that can be considered as conservative estimates.

All the presented results are obtained by means of semi--analytical calculations, while an approach based on numerical simulations is described by Miniati (these proceedings).

\section{Shock waves during structure formation: mergers and accretion}

Clusters of galaxies form hierarchically through the merger of smaller dark matter halos. 
During mergers shock waves naturally form and heat the ICM up to the observed temperatures.

A useful analytical description of this process was proposed by \citet{PS} and later developed by \citet{LC} in the form of the so called extended Press--Schechter formalism. This approach provides us with the comoving number density of clusters with mass $M$ at cosmic time $t$, $n(M,t)$ and with the rate at which clusters of mass $M$ merge at a given time to form clusters with final mass $M^{\prime}$, ${\cal R}(M,M^{\prime},t)$.
It is easy to use this formalism to construct simulated merger trees for clusters with a given present mass.

In \citet{io1} we developed  an analytical recipe that allows us to estimate the Mach number of merger related shocks, once the masses of the two merging clusters  and the redshift at which the merger event occurs are given.
The relative velocity $v_r$ can be evaluated from energy conservation:
\begin{equation}
-\frac{GM_1M_2}{R_1+R_2}+\frac{1}{2}M_rv_r^2 = -\frac{GM_1M_2}{2R_{ta}}
\end{equation}
where $M_i$ and $R_i$ are the masses and the virial radii of the two merging clusters, $M_r$ is the reduced mass and $R_{ta}$ is the turnaround radius, where the two clusters are supposed to be at rest.
The sound speed in the {\it i}-th cluster, needed in order to evaluate the shock Mach number, follows from the virial theorem:
\begin{equation}
c_{s,i}^2 \sim GM_i/(2R_i) \; .
\end{equation}

\begin{figure}
   \centering
   \includegraphics[width=6.5cm]{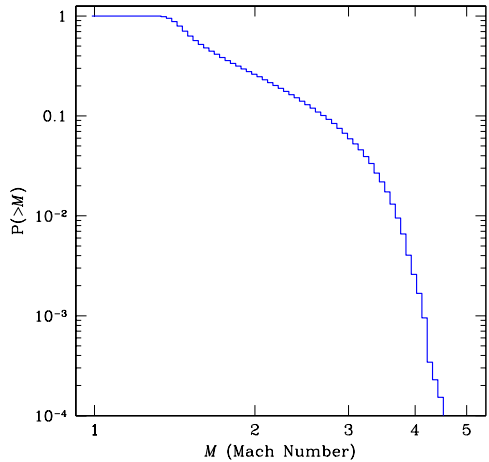}
      
\caption{Cumulative distribution of Mach numbers in mergers of clusters of galaxies \citep{io3}.}
         \label{logmach}
   \end{figure}

The procedure illustrated above can be applied to a generic merger event in the simulated history of a cluster.
The results obtained for 500 different realizations of the merger tree for a cluster with present mass $10^{15} M_{\sun}$ are plotted in fig. \ref{logmach}, where the cumulative probability distribution of shock Mach numbers is shown.

The striking feature of this plot is the fact that most merger related shocks are mildly supersonic, having Mach numbers $\sim 1.5$, while only $6\%$ of the shocks have Mach number greater than $3$ \citep{io1}.

Parallel to the merger processes, a secondary infall of gas occurs at all times onto the potential well which is being formed. The information about the virialization of the inner region of this accretion flow is propagated outwards through a so called accretion shock. While merger related shocks form and propagate in the hot virialized ICM, accretion shocks propagate in the cold ($10^4-10^6 K$), non virialized intergalactic medium, and therefore they are expected to be strong \citep{io2}.

There is still some debate on the typical Mach number of the shocks developed 
in the ICM during cluster mergers. In fact, results obtained by means of numerical simulations suggest the presence of a large fraction of high Mach number shocks in 
cluster mergers \citep{minia}. More recent simulations \citep{ryu}
 seem to find more weak shocks than in \citet{minia}. However the comparison
with the analytical calculations presented here appears difficult 
because of a different classification of the shocks in the two approaches. 
On the other hand, the weakness of the 
merger--related shocks seems to be also suggested by the few observations 
in which the Mach number of the shock can be measured (e.g., \citet{obs}). 

The strength of the shocks is of crucial importance for the acceleration of suprathermal particles and related emission, as discussed in the next section.

\section{Shock acceleration during structure formation}
\label{acc}

Diffusive acceleration can take place at merger and accretion shocks and a small fraction of the particles in the accreting gas is estracted from the thermal distribution and energized up to relativistic energies. 

The maximum energy achievable for a particle can be estimated equating the acceleration time:
\begin{equation}
\tau_{acc} \sim \frac{4D(E)}{u} \; ,
\end{equation}
where $D(E)$ and $u$ are the energy dependent diffusion coefficient and the flow speed, both evaluated upstream \citep{io3}, with the minimum between the energy loss time, the escape time or the life time of the accelerator.
Since we are considering electrons, the relevant time scale is the energy loss time due to ICS off the microwave background radiation photons:
\begin{equation}
\tau_{ICS}=\frac{E}{\frac{4}{3} \sigma_T c u_{cmb} \gamma^2} \; ,
\end{equation}
where $u_{cmb}$ is the energy density in the CMB radiation and the other symbols have their usual meanings.
The highest value of the maximum energy is obtained adopting the Bohm diffusion coefficient:
\begin{equation}
E_{max} \sim 57 \; \left(\frac{B}{\mu G}\right)^{\frac{1}{2}} \; \left(\frac{u}{10^8cm/s}\right) \; TeV \; .
\end{equation}
At these energies the ICS off the CMB photons occurs in the Thomson regime, therefore the maximum energy of the radiated photons is given by:
\begin{equation}
\label{eq.max}
E_{\gamma,max} = \frac{4}{3} \gamma^2 \epsilon_{cmb} 
\end{equation}
$$
\sim 7.5 \; \left(\frac{B}{\mu G}\right) \; \left(\frac{u}{10^8cm/s}\right)^2 \; TeV \; ,
$$
which falls in the very high energy gamma ray band.
It is important to stress, however, that the Bohm diffusion coefficient is the smallest possible one, and that a different choice of the diffusion coefficient could result in a too low maximum energy of the electrons, so that the electron population would be unable to upscatter the CMB photons up to gamma ray energies.Following \citet{l&w} and \cite{io2}, we will use for the maximum energy the value obtained in eq. \ref{eq.max}.

Particle spectra can be easily calculated in the test--particle limit, when the back reaction of the accelerated particles on the shock structure can be neglected. In this case the particle distribution function at suprathermal energies is a power law in momentum $N(p) \propto p^{-\alpha}$ with slope connected to the shock Mach number ${\cal M}$ through the well known relation:
\begin{equation}
\alpha = 2 \; \frac{{\cal M}^2+1}{{\cal M}^2-1}
\end{equation}

It follows that, in order to accelerate flat particle spectra and to have appreciable gamma ray emission strong shocks are needed. As seen in the previous section, accretion shocks are strong, while the majority of merger related shocks are weak and accelerate very steep particle spectra, irrelevant from the point of view of high energy phenomena related to electrons.

\section{Clusters as $\gamma$--ray sources and their contribution to the EDGRB}

The number of accreting clusters detectable by a gamma ray telescope with sensitivity $F_{lim}$ is:
\begin{equation}
N_{acc}(F_{lim}) = 
\end{equation}
$$
\int_0^{\infty} dz \frac{dV}{dz} \int_{M(F_{lim},z)}^{\infty} dM n(M,z)
$$
where $V$ is the comoving volume between redshift $z$ and $z+dz$, $n(M,z)$ is the comoving density of objects as a function of cluster mass and redshift.
The mass of a cluster accreting at redshift $z$ whose flux is $F_{lim}$, $M(F_{lim},z)$, is evaluated assuming that the accretion shock is located at the cluster virial radius and that $5\%$ of the kinetic energy crossing the shock surface is converted into relativistic electrons \citep{io2,io3}.

For merging clusters a similar expression, involving the merger rate ${\cal R}(M,M^{\prime},z)$, holds. The only difference is that for accreting clusters the spectra of the accelerated particles are power laws with fixed slope $\alpha = 2$ while for merging clusters the spectral indexes are evaluated self--consistently as described in \S \ref{acc}.

   \begin{figure}
   \centering
   \includegraphics[width=6.5cm]{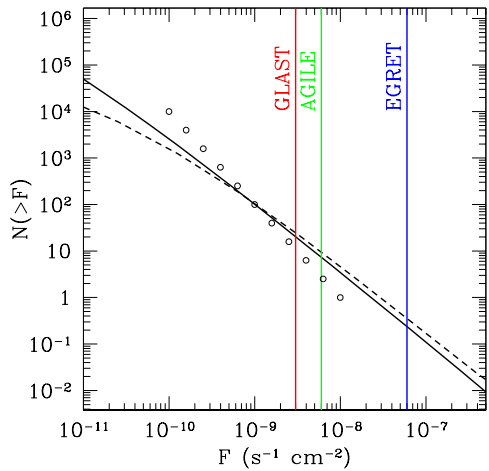}
      
\caption{Number of accreting (solid line) and merging (dashed line) clusters with $\gamma$--ray flux greater than $F$ as estimated by \citet{io3}. Open circles refer to the simulation by \citet{uri}.}
         \label{counts}
   \end{figure}

The number of clusters with gamma ray emission at energies in excess of $100 MeV$ larger that some fixed flux $F$ is plotted in fig. \ref{counts}. The solid and dashed lines refer to accreting and merging clusters respectively. The sensitivity of the EGRET, AGILE and GLAST instruments are also shown.
According to our calculations GLAST and AGILE should detect a few tens and $\sim 10$ clusters respectively. Our results are in agreement with the lack of association between clusters and EGRET sources \citep{olaf}.

\begin{figure}
   \centering
   \includegraphics[width=6.5cm]{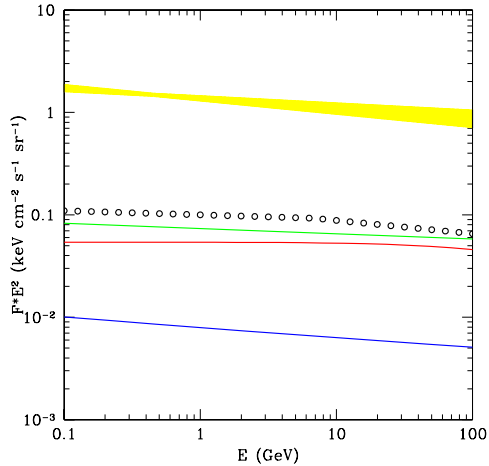}     
\caption{Diffuse gamma ray emission from structure formation (\citet{io2}, see text for details).}
         \label{fondo}
   \end{figure}

The superposition of the emission from single clusters results in a diffuse background. In \citet{io2} we estimated the contribution to this background from forming structure and we summarize our results in fig \ref{fondo}. The shaded yellow area represents the EDGRB as measured by EGRET \citep{egret} while the green and red lines represent the contribution to the background from merging and accreting clusters. A more realistic estimate for the contribution of merging clusters can be obtained assuming a minimum mass for the merging halos of $10^{13} M_{\sun}$ (blue line). 

Since our results greatly differ from some previous results, some 
comments and comparisons are required. \citet{w&l} and \citet{totkita} 
developed
a Press-Schechter based method similar to the one presented here. 
The authors claimed that at least a few tens of clusters should have been 
visible to EGRET. GLAST, on the other hand, was predicted to be able to
detect more than a few thousands of such objects.
This prediction seemed to be supported by some preliminary observational 
evidences for an association between unidentified high latitude 
EGRET sources and Abell clusters (or cluster pairs) presented by \citet{cola} and \citet{kawatot}. The 
statistical significance and physical plausibility of such an association 
was strongly questioned by \citet{olaf}, in which the important conclusion 
was reached that {\it we still have to await the first observational evidence 
for the high-energy gamma-ray emission of galaxy clusters}. 
These findings seem to be
perfectly in line with the predictions described in the present paper. 

All the shocks were assumed to be strong by \citet{totkita} and \citet{w&l}.  As a consequence,
the spectrum of the accelerated particles was always taken to be $\propto
p^{-2}$. As discussed above, and as described in greater details by 
\citet{io1,io2,io3}, for merger related shocks this assumption leads
to incorrect results. The gamma ray emission is 
overestimated by orders of magnitude and its spectrum does not 
reflect the real strength of the shocks developed during mergers of clusters
of galaxies. 
Similar arguments can be used in order to explain the difference between our estimate of the diffuse emission and the results presented by \cite{l&w} and \cite{totkita}, in which the whole EDGRB was saturated by the contribution from structure formation.

Our findings are in good agreement with the results from the numerical simulations by \citet{uri} (open circles in fig. \ref{counts} and \ref{fondo}) in which only gamma rays from ICS were considered and the same assumption we made on acceleration efficiency was used.  

\section{TeV emission from clusters}

\begin{figure}
   \centering
   \includegraphics[width=6.5cm]{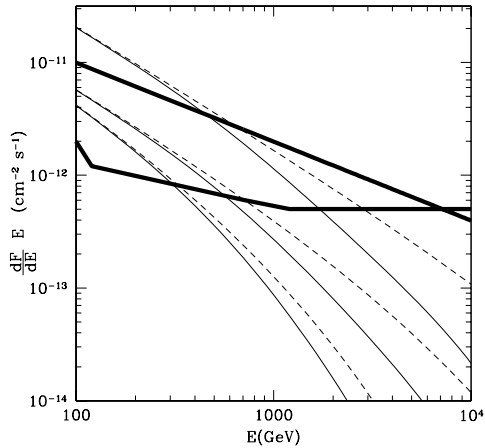}
      
\caption{Gamma ray emission in the TeV region. The thick solid
lines represent the sensitivities of a cherenkov array for point sources (lower
curve) and extended sources (upper curve). The predicted gamma ray fluxes from a Coma-like cluster with and without absorption of the IR--background are plotted as dashed and solid lines.}
         \label{TeV}
   \end{figure}

The maximum energy of the upscattered CMB photons falls in the TeV range (see eq.\ref{eq.max}) so that we can wonder if clusters could be detected by ground based Cherenkov telescopes. In fig. \ref{TeV} we plot the sensitivities for a generic telescope as calculated in \citet{felix} for 100 hours of observation with an
array consisting of 10 cells (pointlike and $1^{\circ}$ sources) together with the expected spectra for three different cases (top to bottom): 1) a merger between two clusters with masses $10^{15}$ and $10^{13} M_{\sun}$, 2) a $10^{15} M_{\sun}$ accreting cluster with a magnetic field in the upstream region equal to
$0.1 \mu G$, 3) the same as 2) but with a ten times lower magnetic field. All the considered clusters are at a distance of $100 Mpc$. The thin continuous and dotted lines refer to the spectra calculated with and without taking into account
the absorption on the infrared background \citep{stecker}.

The apparent size of a $10^{15} M_{\sun}$ accreting cluster at a distance of $100 Mpc$ is comparable or even greater than the field of view of Cherenkov telescopes ($\sim$ a few degrees), while the size of the emitting region for the merger considered in fig. \ref{TeV} should be roughly $1^{\circ}$, making it marginally detectable.

\section{Conclusions}

We developed a recipe to evaluate the Mach numbers of the shocks that form during a merger event \citep{io1}. We found that these shocks are often weak and the corresponding spectra of shock accelerated particles are steep (see also \citet{dermer}). We generalized our approach in \citet{io2,io3} where we considered also the strong accretion shocks and we made predictions about the number of clusters detectable in gamma rays by future gamma ray telescopes and about their contribution to the EDGRB.

According to our prediction, AGILE and GLAST should be able to detect $\sim 10$ and a few tens of objects while the diffuse background produced by clusters is expected to be $\sim 10$ times smaller than that observed by EGRET.

The detection of clusters in the TeV energy range could be problematic due to their extended apparent size.



\bibliographystyle{aa}

\begin{thebibliography}{}

\bibitem[Aharonian et al. (1997)]{felix}
Aharonian, F. A. et al 1997, Astropart. Phys. 6, 369

\bibitem[Berezinsky et al. (1997)]{bbp}
Berezinsky, V. S., Blasi, P. \& Ptuskin, V. S. 1997, ApJ 487, 529

\bibitem[Berrington \& Dermer (2003)]{dermer}
Berrington, R. C. \& Dermer, C. D. 2003, ApJ 594, 709

\bibitem[Blasi (2001)]{pasquale}
Blasi, P. 2001, Astropart. Phys. 15, 223

\bibitem[Colafrancesco (2002)]{cola}
Colafrancesco, S. 2002, A\&A 396, 31

\bibitem[De Jager \& Stecker (2002)]{stecker}
De Jager, O. C. \& Stecker, F. W. 2002, ApJ 566, 738

\bibitem[Feretti,, these proceedings and references therein]{gigia}
Feretti, L. these proceedings

\bibitem[Fusco--Femiano et al. (1999)]{fusco}
Fusco--Femiano, R., et al 1999, ApJ 13, L21

\bibitem[Gabici \& Blasi (2003a)]{io1}
Gabici, S. \& Blasi, P. 2003a, ApJ 583, 695

\bibitem[Gabici \& Blasi (2003b)]{io2}
Gabici, S. \& Blasi, P. 2003b, Astropart. Phys. 19, 679

\bibitem[Gabici \& Blasi (2004)]{io3}
Gabici, S. \& Blasi, P. 2004, Astropart. Phys. in press (astro-ph/0306369)

\bibitem[Kawasaky \& Totani (2000)]{kawatot}
Kawasaky, W. \& Totani, T. 2000, ApJ 576, 679

\bibitem[Keshet et al. (2003)]{uri}
Keshet, U., Waxman, E. \& Loeb, A. 2003, ApJ 585, 128

\bibitem[Lacey \& Cole (1993)]{LC}
Lacey, C. \& Cole, S. 1993, MNRAS 262, 627

\bibitem[Loeb \& Waxman (2000)]{l&w}
Loeb, A. \& Waxman, E. 2000, Nature 405, 156

\bibitem[Markevitch et al. (2003)]{obs}  
Markevitch M. et al 2003,
in 'Matter and Energy in Clusters
 of Galaxis', ASP Conf. Series, 301, p.37,
eds. S. Bowyer and C.-Y. Hwang

\bibitem[Miniati, these proceedings]{miniathis}
Miniati, F. these proceedings

\bibitem[Miniati et al. (2000)]{minia}
Miniati, F. et al 2000, ApJ 542, 608

\bibitem[Press \& Schechter (1974)]{PS}
Press, W. H. \& Schechter, P. 1974, ApJ 187, 425

\bibitem[Reimer et al. (2003)]{olaf}
Reimer, O. et al 2003, ApJ 588, 155

\bibitem[Ryu et al. (2003)]{ryu}
Ryu, D. et al 2003, ApJ 593, 599

\bibitem[Sreekumar et al. (1998)]{egret}
Sreekumar, P. et al. 1998, ApJ 494, 523

\bibitem[Totani \& Kitayama (2000)]{totkita}
Totani, T. \& Kitayama, T. 2000, ApJ 545, 572

\bibitem[Voelk et al. (1996)]{vab}
Voelk, H. J. et al. 1996, Space Sci. Rev. 75, 279

\bibitem[Waxman \& Loeb (2000)]{w&l}
Waxman, E. \& Loeb, A. 2000, ApJ 545, L11

\end{thebibliography}

\end{document}